\documentclass[sigconf]{acmart}
\AtBeginDocument{%
  }

\usepackage{booktabs}

\usepackage{array}   

\setcopyright{none}
\renewcommand\footnotetextcopyrightpermission[1]{}

\acmConference[arXiv~'26]{arXiv}{September}{2026}

\begin{document}

\title{Sona: Personalized Soundscape Mediation to Support People with Sound Sensitivity} 

\author{Jeremy Zhengqi Huang}
\affiliation{%
    \institution{University of Michigan}
    \city{Ann Arbor, MI}
    \country{USA}}
\email{zjhuang@umich.edu}

\author{Emani Hicks}
\affiliation{%
    \institution{University of Houston}
    \city{Houston, TX}
    \country{USA}}
\email{eihicks2@central.uh.edu}

\author{Sidharth}
\affiliation{%
    \institution{University of Michigan}
    \city{Ann Arbor, MI}
    \country{USA}}
\email{sidcs@umich.edu}

\author{Gillian R. Hayes}
\affiliation{%
    \institution{University of California, Irvine}
    \city{Irvine, CA}
    \country{USA}}
\email{gillianrh@ics.uci.edu}

\author{Dhruv Jain}
\affiliation{%
    \institution{University of Michigan}
    \city{Ann Arbor, MI}
    \country{USA}}
\email{profdj@umich.edu}

\renewcommand{\shortauthors}{Jeremy Zhengqi Huang, Emani Hicks, Sidharth, Gillian R. Hayes, Dhruv Jain}

\begin{abstract}
People with sound sensitivity (PWSS) often manage distressing sounds with earplugs and noise-canceling headphones that broadly suppress their surroundings, limiting access to useful auditory cues. We present Sona, a mobile system for personalized, real-time soundscape mediation, informed by prior sound sensitivity research and an online survey of 68 PWSS. Sona selectively attenuates multiple overlapping user-chosen sounds at adjustable strength, suggests targets from ambient sound recognition, and lets users add custom targets from short recordings without retraining the model. In an in-situ evaluation with ten PWSS, participants reported that Sona made their soundscapes more manageable. The study also surfaced uneven attenuation across sound type and context, tensions between managing filters and attending to ongoing activities, and difficulty interpreting personalization outcomes. These findings highlight the need to design for the quality of the residual soundscape, balance user control with interaction demands, and support guided, interpretable personalization.

\end{abstract}



\begin{teaserfigure}
  \centering
  \includegraphics[width=\textwidth]{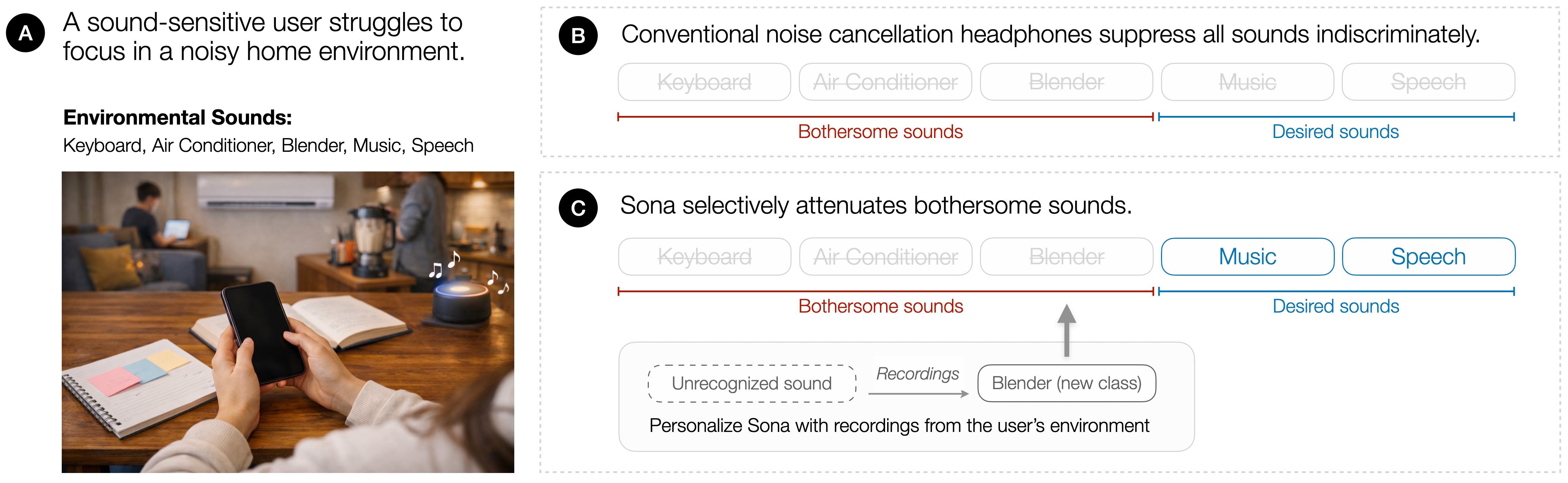}
  \caption{Sona helps people with sound sensitivity manage challenging everyday soundscapes. (A) In a shared home, overlapping sounds such as keyboard typing, air conditioner hum, blender noise, and music can make it difficult to focus. (B) Conventional noise-canceling headphones broadly suppress the soundscape, reducing both bothersome and desired sounds. (C) Sona instead selectively attenuates bothersome sounds. Users can also personalize Sona by adding new sound targets from recordings of their environments.}
  \Description{A three-part overview figure illustrating Sona, a system for helping people with noise sensitivity manage everyday acoustic environments. Panel A shows a user at a table in a shared home setting with a laptop user in the background, a blender in the kitchen, an air conditioner on the wall, and a speaker playing music. The panel text says the user struggles to focus in a noisy home environment, with environmental sounds including keyboard, air conditioner, blender, and music. Panel B shows conventional noise-canceling headphones treating all sounds the same: keyboard, air conditioner, blender, music, and speech are all muted, with no distinction between bothersome and desired sounds. Panel C shows Sona selectively attenuating bothersome sounds while preserving desired sounds. Keyboard, air conditioner, and blender are shown as bothersome sounds, while music and speech are preserved as desired sounds. A smaller subpanel below shows personalization over time: an unrecognized sound can become a new sound class, labeled “Blender,” allowing the system to be extended with sounds from the user’s own environment.}
  \label{fig:teaser}
\end{teaserfigure}

\received{20 February 2007}
\received[revised]{12 March 2009}
\received[accepted]{5 June 2009}

\settopmatter{printacmref=false}

\maketitle

\section{Introduction}

Everyday soundscapes can be both resources and barriers: they support communication, awareness, and safety; yet routine sounds can cause distress for people with sound sensitivity (PWSS). In this paper, we use PWSS as an experience-based term for diverse forms of decreased sound tolerance, including misophonia---intense reactions to particular everyday sounds \cite{swedo_consensus_2022}---and hyperacusis, in which sounds others tolerate feel excessively loud or painful \cite{noauthor_what_2026}. These experiences can disrupt concentration and participation across home, work, and public life. Their reach is substantial: nearly one in twenty U.S. adults report clinically significant misophonia \cite{dixon_prevalence_2024}, and a meta-analysis of 67 studies estimated that hyperacusis affects 41.4\% of autistic people currently and 60.6\% at some point in their lives \cite{williams_review_2021}.

Existing tools, such as noise-canceling headphones, earplugs, or audio masking (e.g., white noise), help PWSS limit trigger exposure by broadly blocking or masking environmental sound. However, these broad, fixed controls cannot accommodate either the personal and contextual nature of sound sensitivity or the unpredictability of real-world trigger exposure \cite{dotch_understanding_2023, hicks_informing_2025}. For instance, listeners must often choose between blocking or masking sound in advance, which reduces access to speech, alarms, and other useful cues even when no trigger occurs, or reacting only after a trigger has already occurred. This tradeoff demands constant vigilance while giving PWSS little control over which sounds are reduced and which remain accessible.

Recent advances in audio machine learning (ML), including classification \cite{noauthor_sound_2026}, audio language models \cite{ghosh_audio_2026}, and source separation \cite{liu_separate_2024, veluri_real-time_2023, shi_sam_2025}, present opportunities for a more nuanced approach to managing trigger exposure. Prior systems, for example, used target sound extraction capabilities to selectively enhance or suppress target sounds and speech in real time \cite{veluri_real-time_2023, veluri_semantic_2023, oh_fine-grained_2026, veluri_look_2024}. However, translating these capabilities into support for PWSS raises an open question: how to design soundscape mediation systems that accommodate the personal and situated nature of sound sensitivity \cite{hicks_informing_2025, dotch_understanding_2023}?

To explore this question, we synthesized prior work and conducted an online survey with 68 participants who self-reported sound sensitivity. Participants reported variation in the sounds they found bothersome and in how strongly they wanted those sounds changed. However, reduction and removal were the most commonly preferred interventions. Combining these findings with prior work, we derived three design goals: support targeted, graduated control over bothersome sounds; minimize in-the-moment interaction burden while preserving user control; and support personalization beyond a fixed set of sound classes.


Guided by these considerations, we developed Sona, a mobile system that helps PWSS mediate their real-time soundscape. Sona extends real-time target sound extraction \cite{veluri_real-time_2023, veluri_semantic_2023} with a multi-target neural attenuation framework that fuses embeddings of selected sounds to attenuate multiple overlapping triggers in a single on-device inference pass, allowing users to selectively attenuate multiple overlapping triggers at an adjustable strength. Users can select targets manually or accept suggestions from a sound recognizer that actively detects sounds in the environment. Sona supports 25 built-in sound classes informed by our formative inquiry and lets users add custom targets from recordings \emph{without retraining the model}. To help capture relevant recordings, users can describe their triggers in natural language, such as ``mechanical whirring'' or ``repetitive tapping.'' Sona matches these sensitivities against descriptions of ambient audio generated by an audio language model \cite{goel_audio_2025} and prompts users to save relevant recordings for personalization. Sona runs in real time on a commodity smartphone (approximately 42~ms end-to-end latency) and, in benchmark mixtures, attenuates up to three simultaneous targets with stable improvement over the unprocessed mixture.


To understand how PWSS experience and personalize Sona, we conducted a user study with ten PWSS in two-hour sessions. Each session included \emph{in-situ} listening activities in which participants used Sona in a study lounge and a busy food court, followed by a personalization task and a semi-structured interview. Participants appreciated when Sona reduced bothersome sounds while other environmental cues remained audible, giving them a sense of control over the soundscape. However, they reported uneven attenuation across sound types and contexts and described tensions between controlling the soundscape and sustaining attention to ongoing activities. Participants valued the ability to personalize the system with custom targets but wanted clearer guidance and feedback during the process. These findings point to the importance of designing for the quality of the remaining soundscape, reducing interaction demands while preserving user control, and supporting guided, interpretable personalization.



In summary, our work makes two contributions to HCI research on accessible soundscape technologies. First, we present Sona, a mobile system that builds on real-time selective sound extraction \cite{veluri_real-time_2023, veluri_semantic_2023, oh_fine-grained_2026} to support the needs of PWSS documented in prior accessibility research \cite{hicks_informing_2025, dotch_understanding_2023}. Guided by formative inquiry, Sona combines suggestions during trigger encounters, adjustable multi-target attenuation, and custom sound class enrollment from short recordings without retraining. Second, we contribute empirical knowledge about how people with sound sensitivity experience real-time soundscape mediation in real-world settings. We uncover tensions between exercising soundscape control and sustaining attention to ongoing activities, illuminate challenges in specifying personal triggers and assessing personalization outcomes, and derive design implications that extend the HCI community's understanding of what it means to design for dynamic, personalized auditory accessibility.

\section{Related Work}

Our research sits at the intersection of four areas: (1) lived experiences of sound sensitivity, (2) interactive soundscape mediation, (3) audio ML for sound separation, and (4) end-user personalization in assistive technology. Together, this literature motivates real-time selective mediation that is grounded in the needs of PWSS.

\subsection{Experiences and Needs in Sound Sensitivity}
Prior work shows that sound sensitivity is not simply a function of loudness, but a highly situated and individualized experience shaped by sound characteristics, context, predictability, and emotional state \cite{dotch_understanding_2023, hicks_informing_2025}. Everyday sound can become a continuous interactional demand that must be monitored and managed across daily life, influencing where people work, socialize, and travel \cite{dotch_understanding_2023, henry_sound_2022}. Managing auditory overload is closely tied to emotional regulation and participation in everyday activities \cite{das_towards_2021, kim_workplace_2022, hall_designing_2024}.

Existing coping strategies, such as leaving the environment, covering ears, or using earplugs and noise-canceling headphones, can provide relief but often reduce access to environmental and social information \cite{dotch_understanding_2023, hicks_informing_2025}. This tension between \emph{comfort} and \emph{awareness} also appears in HCI work on soundscape curation and hear-through listening which shows 
that people do not simply want to block sound, but to negotiate competing needs for relief, awareness, attention, and social participation \cite{haas_cant_2018, johansen_characterising_2022, johansen_personalised_2019}. 

Crucially, what counts as bothersome varies substantially across individuals and contexts. Misophonia, autism-related sensitivities, and hyperacusis each involve different triggers and mechanisms \cite{simner_automated_2024, brout_investigating_2018, williams_review_2021, aazh_sound_2024}, and even within individuals, responses to sound can shift with context and emotional state \cite{brout_investigating_2018, dotch_understanding_2023}. Together, this literature motivates systems that support \emph{selective, personalized, and context-aware} sound mediation rather than one-size-fits-all suppression.

\subsection{Soundscape Mediation Tools}
Research in HCI has increasingly framed soundscape perception as an interactive activity rather than as a purely acoustic phenomenon. Early work showed that people already engage in situated sound-management practices in domestic environments \cite{oleksik_sonic_2008}. More recent work conceptualizes soundscapes both as \emph{acoustic environments} and as \emph{compositions} that can be curated and modified through interactive systems \cite{johansen_characterising_2022, haas_interactive_2020, chang_soundshift_2024, cao_soundmodvr_2024}. This line of work establishes that headphones and hear-through systems are not merely tools for isolation, but interfaces for negotiating comfort, awareness, and social acceptability of real-life auditory environments \cite{haas_cant_2018}. More recent systems---such as Semantic Hearing \cite{veluri_semantic_2023}---pushed the trajectory toward selective or programmable soundscapes that allow users to foreground or suppress specific sound classes. However, these systems were not specifically designed for or evaluated with PWSS, whose sound-management needs vary across individuals, activities, and environments. This work builds on this trajectory by investigating how real-time soundscape mediation can support PWSS, grounding its design in their experiences and evaluating its use in situ.


\subsection{Audio ML for Sound Modification}
Advances in audio ML have made it increasingly feasible to selectively reshape acoustic scenes rather than merely reduce noise globally. Early neural separation and enhancement models, such as Conv-TasNet \cite{luo_conv-tasnet_2019}, Demucs \cite{defossez_music_2021}, and DCCRN \cite{hu_dccrn_2020}, established that learned models can perform high-quality source separation and denoising with latencies approaching interactive use. 

Building on this foundation, later work shifted toward target-conditioned and semantically flexible control. Universal sound separation systems and arbitrary-sound extraction approaches, such as SoundFilter \cite{gfeller_one-shot_2020}, showed that a single model could isolate a wide range of sounds or follow a short reference example. Subsequent work extended this paradigm to higher-level control: CLIPSep \cite{dong_clipsep_2023} and AudioSep \cite{liu_separate_2024} demonstrated that natural-language descriptions can specify which sounds to extract from a mixture, enabling more flexible and user-steerable interaction. More recently, SAM Audio \cite{shi_sam_2025} further generalized this direction by treating audio separation as a promptable task across text, visual, and temporal-span inputs. Together, this line of work shifts audio processing from fixed denoising pipelines toward \emph{user-steerable sound manipulation}, where a single model can be conditioned to produce different outputs. Most closely related to Sona is Semantic Hearing \cite{veluri_semantic_2023}, which brings target-conditioned audio separation into real-time listening contexts by allowing users to foreground or suppress specified sound sources while preserving environmental awareness. However, prior real-time target sound extraction systems for environmental sound, including Semantic Hearing \cite{veluri_semantic_2023}, operate over a \emph{fixed vocabulary} of sound classes, and extending them to new sounds typically requires \emph{retraining} or \emph {predefined labels}. Sona extends this line of work with a unified framework that keeps the attenuation model fixed and on-device while allowing the set of target classes to grow from user recordings.


\subsection{End-User Personalization in Accessibility}
Accessibility research emphasizes that assistive technologies must adapt to highly individualized and evolving user needs. Prior work shows that users often customize, combine, or repurpose tools to address long-tail accessibility requirements \cite{hurst_making_2013, herskovitz_hacking_2023}. This is especially important in sound sensitivity, where bothersome sounds vary widely across individuals and contexts \cite{brout_investigating_2018, scheerer_autistic_2024, hicks_informing_2025}. More broadly, prior work argues that end-user-trained ML systems can be especially effective when the learning task is constrained to a specific user and their environment, achieving a greater real-world robustness than generic pretrained models alone \cite{kacorri_teachable_2017}.

This paradigm has been explored most extensively in the audio domain through sound-awareness tools \cite{jain_protosound_2022, goodman_toward_2021}, which allow users to train recognition models to detect sounds relevant to them. In contrast, Sona enables users to teach new \textit{suppression} targets, extending the system’s vocabulary through example-driven workflows grounded in their own environments. By combining end-user personalization with target-conditioned audio models, Sona is designed to evolve with use.

\section{Formative Study}
To ground the design of the sound-management tool in the experiences of people with sound sensitivity, we conducted an online survey with 68 participants. The survey focused on two questions: which everyday sounds participants most wanted to manage, and how they preferred those sounds to be handled. 

\noindentparagraph{\textbf{Method:}} 
We first reviewed studies documenting the auditory experiences of sound-sensitive populations, including work on misophonia, hyperacusis, and sensory sensitivities in autism \cite{scheerer_autistic_2024, simner_automated_2024, vitoratou_listening_2021, jastreboff_decreased_2015}, and extracted 59 everyday sounds that were repeatedly cited as problematic. These were organized into 12 categories (\textit{e.g.,} appliances, tools, traffic, loud impact sounds, media).

We then designed an online survey to assess which sounds from this inventory participants found bothersome and how they preferred to manage them. The survey consisted of four blocks: (1) a validated screener for sound sensitivity (SSSQ-2) \cite{aazh_sound_2024}, (2) category-level and item-level bothersomeness ratings, (3) preferred management strategies for bothersome sounds, and (4) demographics. For each category, participants first indicated whether it was bothersome in everyday life. If so, they selected specific sound items within that category and indicated how they would prefer to handle them. Response options included leaving the sound unchanged, lowering or reducing it, masking it with pleasant sounds, modifying its tone or style, or removing it entirely. Our institution's Institutional Review Board (IRB) determined that the study qualified for Exemption 2 status.

We received 360 responses. Eligibility required being at least 18 years old, residing in the United States, and reporting at least one SSSQ-2 symptom on two or more days during the preceding two weeks; because the SSSQ-2 does not define a validated cutoff, this frequency threshold was a study-specific criterion. After excluding ineligible respondents and responses that showed signs of fradulent or automated submission (e.g., auto-generated email addresses, duplicate entries) or careless responding, we retained 68 respondents (19\%).

SSSQ-2 total scores averaged 7.01 out of 18 ($SD = 3.47$, range = 1--17). For context, the instrument's validation sample, recruited through university students and staff and tinnitus and sound-intolerance support groups, had a mean of 6.6 ($SD = 4.3$)~\cite{aazh_sound_2024}. Following the instrument's item structure, mean scores were 4.28 ($SD = 2.38$) for the combined hyperacusis items (0--12), 1.26 ($SD = 1.06$) for the misophonia item (0--3), and 1.47 ($SD = 1.04$) for the noise sensitivity item (0--3). These scores describe self-reported symptom frequency rather than clinical diagnoses. Among the 64 respondents who provided demographic information, 12 were aged 18–24, 32 were 25–34, 14 were 35–44, and 6 were 45 or older. For gender, 28 identified as women, 26 as men, 6 as nonbinary, and 2 as transgender; 2 preferred not to disclose their gender.


\subsection{Survey Findings}

\begin{figure}[t]
    \centering
    \includegraphics[width=0.8\linewidth]{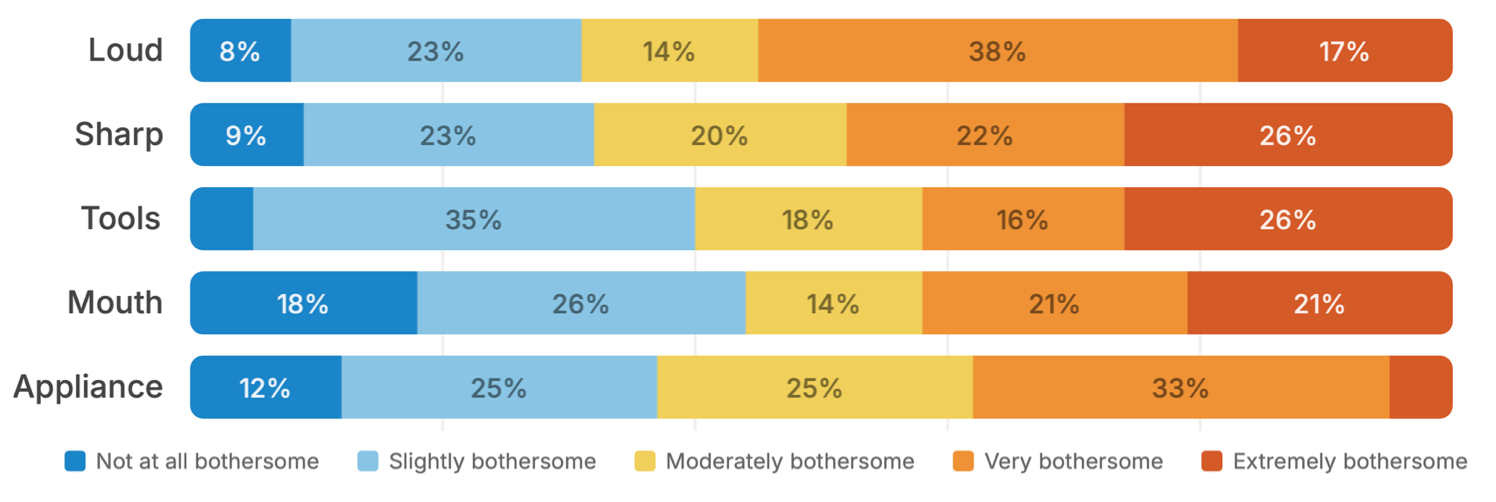}
    \caption{Distribution of participants' reported bothersomeness across sound categories (Top 5, sorted by ratings of very bothersome or above).}
    \Description{A horizontal stacked bar chart showing the distribution of bothersomeness ratings across five sound categories. Each bar is divided into five segments: Not at all bothersome (dark blue), Slightly bothersome (light blue), Moderately bothersome (yellow), Very bothersome (orange), and Extremely bothersome (dark orange). Loud sounds: 8\% not at all, 23\% slightly, 14\% moderately, 38\% very, 17\% extremely. Sharp sounds: 9\% not at all, 23\% slightly, 20\% moderately, 22\% very, 26\% extremely. Tools: 35\% slightly, 18\% moderately, 16\% very, 26\% extremely. Mouth sounds: 18\% not at all, 26\% slightly, 14\% moderately, 21\% very, 21\% extremely. Appliance sounds: 12\% not at all, 25\% slightly, 25\% moderately, 33\% very. Loud and Sharp categories show the highest combined proportions of very and extremely bothersome ratings.}
    \label{fig:survey_result}
    \vspace{-1em}
\end{figure}

\noindentparagraph{\textbf{Finding 1: Bothersome sounds are highly individualized, but certain categories are consistently salient.}}
Participants reported substantial variability in which sounds were bothersome, reinforcing prior work on the individualized nature of noise sensitivity. However, some categories emerged as consistently problematic. Loud and sharp sounds were most frequently rated as very or extremely bothersome, followed by tools and mouth sounds (Figure~\ref{fig:survey_result}). Within these categories, sounds such as lawn mowers, leaf blowers, vacuum cleaners, metal scraping, and chewing-related sounds were identified as especially bothersome. These results suggest that while broad categories can guide system design, the specific sounds that matter remain highly user-dependent.

\noindentparagraph{\textbf{Finding 2: Preferred sound-management strategies vary by sound category and annoyance level, but reduction and removal dominate.}}
Across categories, participants most often preferred lowering or reducing unwanted sounds, followed by completely removing them (Table~\ref{tab:sound_preferences}). Fewer respondents preferred modifying the tone or style of sounds, masking with pleasant sounds, or leaving them unchanged. These preferences varied by sound type. For voices and mouth sounds, respondents preferred reduction that preserved some awareness, whereas for taps, alarms, and sharp sounds, they preferred complete removal. Preferences also shifted with perceived annoyance, with higher annoyance associated with stronger interventions, particularly complete removal. These results indicate that while strategies vary with sound type and perceived annoyance, reduction and removal consistently dominate. 

\begin{table}[t]
    \caption{Participant preferences for modifying unwanted sounds (multi-selection allowed).}
    \label{tab:sound_preferences}
    \vspace{-1em}
    \centering
    \begin{tabular}{lr}
        \toprule
        \textbf{Modification Preference} & \textbf{Percentage} \\
        \midrule
        \textbf{Lower / reduce the sound} & \textbf{36.2\%} \\
        \textbf{Completely remove the sound} & \textbf{29.1\%} \\
        Change the tone and style & 16.5\% \\
        Play pleasant sounds on top & 15.3\% \\
        Leave them as is & 13.7\% \\
        \bottomrule
    \end{tabular}
    \vspace{-1em}
\end{table}

\subsection{Design Goals}
\label{sec:design_goals}
Combining prior work (Section 2) with these findings, we derived three design goals for sound-sensitivity support technology.

\noindentparagraph{\textbf{DG1: Support targeted and graduated control over bothersome sounds.}}
People with sound sensitivity often seek relief without losing awareness of their surroundings. Our survey shows that preferred interventions range from reduction to removal depending on the sound type and perceived annoyance. This motivates selectively targeting specific sounds and adjusting attenuation strength, rather than applying uniform suppression.

\noindentparagraph{\textbf{DG2: Minimize in-the-moment interaction burden while preserving user control.}}
Prior work shows that moments of sensory overload can limit users' ability to actively manage sound \cite{dotch_understanding_2023, hicks_informing_2025}. This motivates lightweight interactions and just-in-time assistance that reduce the need to anticipate every sound while still preserving user agency.

\noindentparagraph{\textbf{DG3: Support personalization beyond a fixed sound inventory.}}
Both prior work \cite{hicks_informing_2025, landon_qualitative_2016} and our survey results highlight substantial variation in which sounds are bothersome and how they are experienced. This motivates systems that can be extended with user-specific sounds over time, rather than relying on a fixed inventory.

\section{Sona}
\label{sec:sona_system}

\begin{figure*}[ht]
    \centering
    \includegraphics[width=\linewidth]{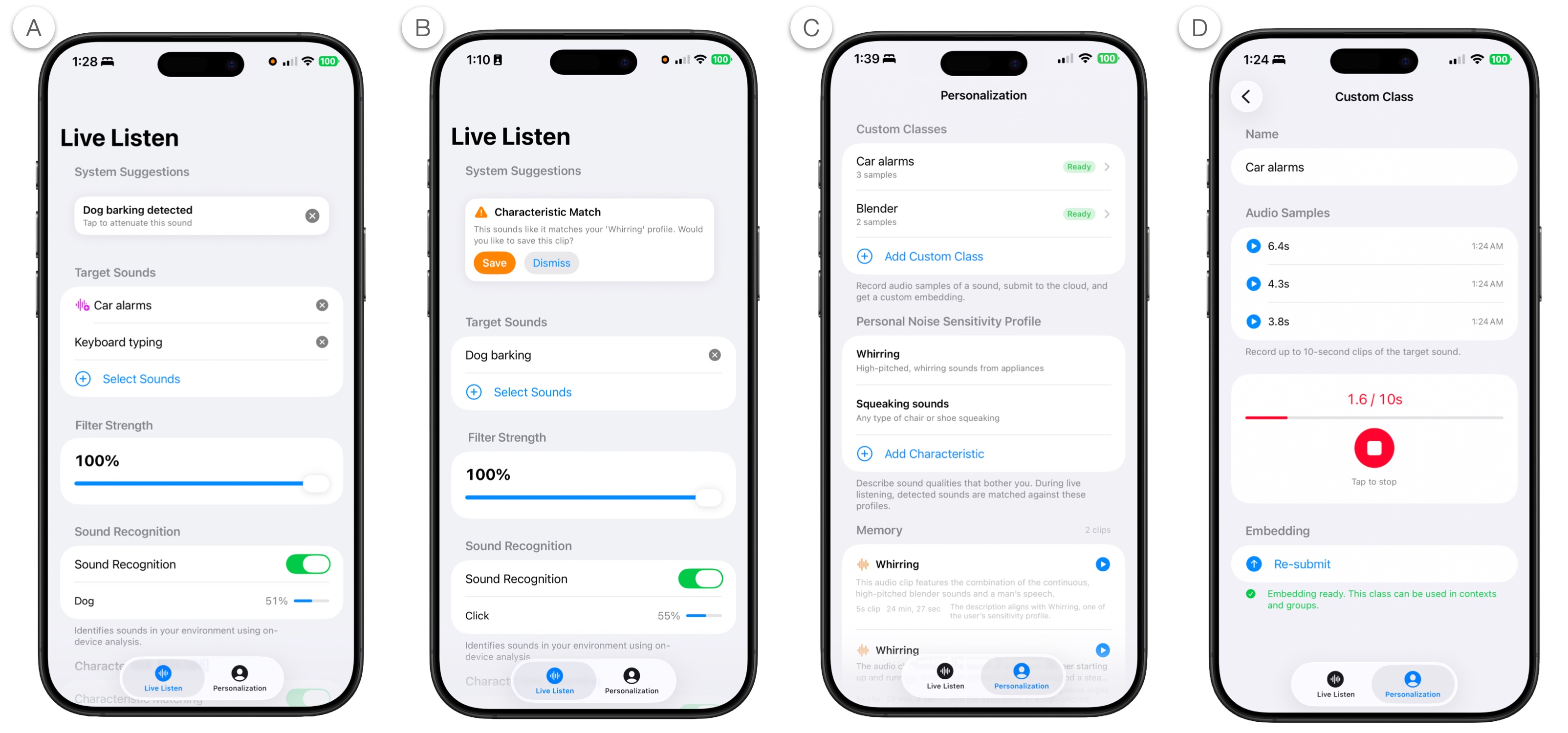}
    \vspace{-1em}
    \caption{Sona's User Interface. (A) In Live Listen, Sona surfaces suggestions for detected sounds, lets users select target sounds, and supports real-time adjustment of attenuation strength. (B) When an unfamiliar sound matches a stored sensitivity profile, Sona prompts the user to save it for later personalization. (C) Users can manage custom sound classes and sensitivity profiles. (D) Users can create a reusable custom class by recording audio samples.}
    \Description{Four iPhone screenshots showing Sona's user interface. Panel A shows the Live Listen view with a system suggestion banner reading "Dog barking detected, Tap to attenuate this sound," two active target sounds (Car alarms and Keyboard typing) each with a remove button, a Filter Strength slider set to 100\%, and a Sound Recognition section showing "Dog" detected at 51\% confidence. Panel B shows the Live Listen view with a Characteristic Match alert warning that a sound matches the user's "Whirring" sensitivity profile, with Save and Dismiss buttons. One target sound (Dog barking) is active, Filter Strength is at 100\%, and Sound Recognition shows "Click" at 55\% confidence. Panel C shows the Personalization view with two sections: Custom Classes listing "Car alarms" (3 samples, Ready) and "Blender" (2 samples, Ready) with an Add Custom Class button, and Personal Noise Sensitivity Profiles listing "Whirring" described as "High-pitched, whirring sounds from appliances" and "Squeaking sounds" described as "Any type of chair or shoe squeaking," with an Add Characteristic button. A Memory section below shows two saved audio clips labeled "Whirring" with natural-language descriptions and timestamps. Panel D shows the Custom Class creation view for "Car alarms" with three recorded audio samples (6.4s, 4.3s, and 3.8s), a recording interface showing 1.6 of 10 seconds elapsed with a red stop button, and an Embedding section showing a green checkmark indicating the embedding is ready for use.}
    \label{fig:user_interface}
    \vspace{-1em}
\end{figure*}

\begin{figure*}[ht]
    \centering
    \includegraphics[width=\linewidth]{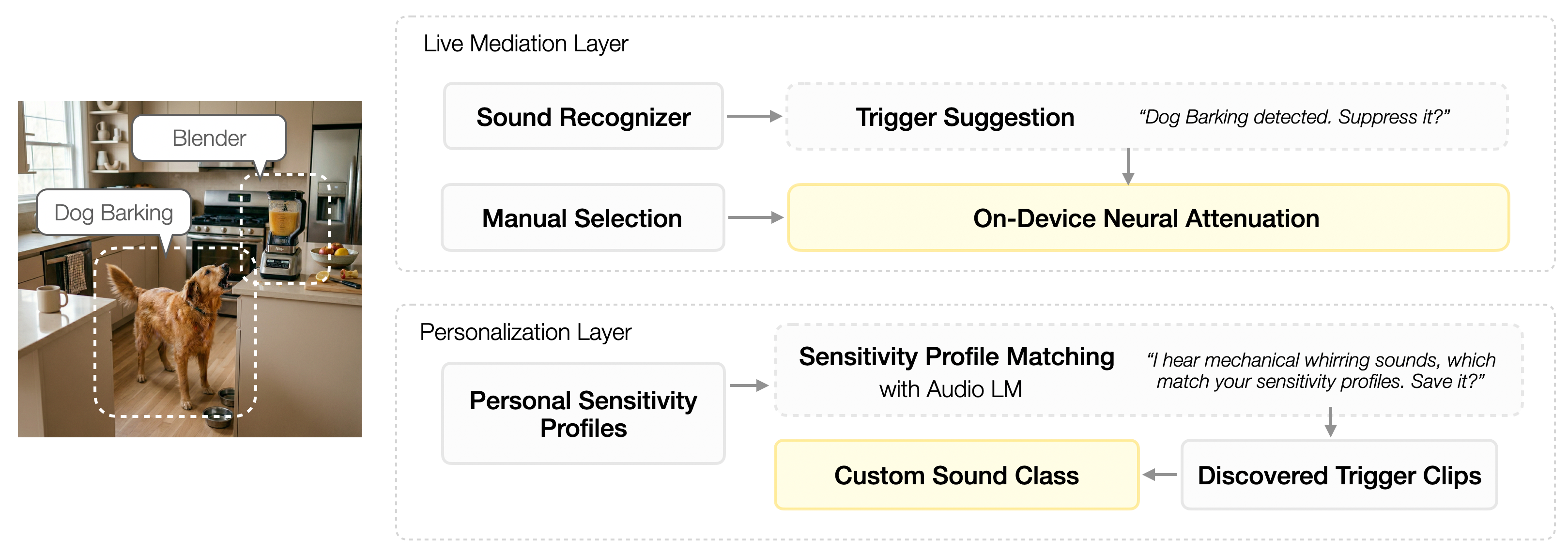}
    \caption{Sona's two-layer system architecture. The Live Mediation Layer runs on-device: users select target sounds manually or accept suggestions from a sound recognizer, and the neural attenuation model reduces the selected sounds at an adjustable strength while preserving the rest of the scene. The Personalization Layer extends the built-in sound set: users describe personal triggers as sensitivity profiles, Sona matches audio-language-model descriptions of the live soundscape against these profiles and prompts users to save matching clips, and users create custom sound classes from saved or manually recorded examples.}
    \Description{Two-layer diagram beside a kitchen photo in which a barking dog is labeled "Dog Barking" and a running blender is labeled "Blender." In the Live Mediation Layer, a Sound Recognizer box points to a Trigger Suggestion box with the example prompt "Dog Barking detected. Suppress it?" Both the Trigger Suggestion and Manual Selection boxes point to an On-Device Neural Attenuation box. In the Personalization Layer, a Personal Sensitivity Profiles box points to a Sensitivity Profile Matching with Audio LM box with the example prompt "I hear mechanical whirring sounds, which match your sensitivity profiles. Save it?"; it points down to Discovered Trigger Clips, which points to Custom Sound Class.}
    \label{fig:system_architecture}
    \vspace{-1em}
\end{figure*}

Sona is a mobile system that helps people with sound sensitivity mediate their live soundscapes. In this section, we describe the underlying multi-target audio attenuation framework that powers Sona (Section~\ref{sec:attenuation_framework}) and the system architecture (Figure~\ref{fig:system_architecture}) that works towards the three design goals identified in the previous section. Figure~\ref{fig:user_interface} shows the corresponding interfaces. 

\subsection{Multi-Target Neural Attenuation Framework}
\label{sec:attenuation_framework}
Our framework supports the identified design goals by enabling targeted control of multiple simultaneous bothersome sounds (Section~\ref{sec:sona-multi-target}) and end-user extension of sound classes without retraining the model (Section~\ref{sec:sona-class-extension}).


\subsubsection{Supporting Multi-Target Attenuation}
\label{sec:sona-multi-target}
In an office, a user may feel overwhelmed by keyboard typing, HVAC hum, and intermittent construction noise, and wish to attenuate them all while hearing other sounds. Addressing this requires the model to represent and process several targets within a smartphone's computational budget. Therefore, our framework implements attenuation using an adapted Deep Complex Convolutional Recurrent Network (DCCRN) \cite{hu_dccrn_2020}, which receives the microphone mixture and embeddings representing the selected sounds. The network then estimates the soundscape that should remain after attenuation. To compose the selected targets, our framework adapts the embedding-fusion approach from MTSE \cite{serre_mtse_2025}. For the $K$ selected embeddings, the module computes
\begin{equation}
\mathbf{e}_{\mathrm{fused}} = W_2\!\left[
\max_{k=1,\ldots,K}\operatorname{SiLU}(W_1\mathbf{e}_k+b_1)
\right]+b_2.
\label{eq:sona-target-fusion}
\end{equation}
Learned projections and element-wise maximum pooling fuse the selected embeddings into a fixed-dimensional vector. This vector conditions the DCCRN bottleneck through feature-wise linear modulation (FiLM) \cite{perez_film_2017}, allowing all selected targets to be processed in a single forward pass.

\subsubsection{Making the Sound Class Extensible}
\label{sec:sona-class-extension}
Prior real-time target extraction systems conditioned extraction on class-label vectors defined over a pre-defined vocabulary of sound categories \cite{oh_fine-grained_2026, veluri_semantic_2023}. These systems support selecting known categories but do not provide an end-user mechanism for extending the system's coverage. A predefined catalog may not adequately represent a user's particular triggers, such as a bathroom fan's whine, an upstairs treadmill's thud, or an office chair's squeak. Requiring model retraining for each new target would make customization difficult to sustain as users encounter additional triggers. Therefore, we designed the framework so that personalization happens at the level of \emph{target representation}, not network weights. Sona maintains an external embedding store:
\[
\mathcal{D} = \{(c_i, \mathbf{v}_i)\}_{i=1}^{M},
\]
which maps a class identifier $c_i$ to a 768-dimensional embedding $\mathbf{v}_i$. Built-in sound classes ship with precomputed embeddings. User-created sound classes are generated upstream from short recordings collected in the user's own environment and inserted into the same store once their embeddings are ready.

This makes end-user extension of sound classes operationally simple, as built-in and future sound classes share the same embedding store, fusion module, and attenuation model. With this design, adding a personal sound therefore requires a compatible embedding, without retraining or replacing the deployed model.

\subsubsection{Model Training and Target Embeddings}
\label{sec:sona-training}
\paragraph{Training data.}
We construct synthetic mixtures from VGGSound \cite{chen_vggsound_2020}, ESC-50 \cite{piczak_esc_2015}, and FSD50K \cite{fonseca_fsd50k_2022}, covering 25 target categories informed by the formative survey. Mixtures combine four-second recordings resampled to 16\,kHz. We designate a subset of sources for attenuation, retain at least one source, and scale each target to a signal-to-interference ratio sampled uniformly from 0 to 10\,dB relative to the retained sources. The reference output is the sum of the unselected sources, making their reconstruction part of the objective alongside removal of the selected sounds.

\paragraph{Target embeddings.}
We use the same pretrained AudioSep representation pipeline \cite{liu_separate_2024} for built-in and personal classes. For each recording, we extract intermediate encoder activations near the separation backbone's bottleneck, collapse the frequency dimension, and apply attentive statistics pooling over time \cite{okabe_attentive_2018}. This converts variable-duration recordings into fixed-dimensional representations, which we average across a class's examples to obtain its target embedding.

\paragraph{Attenuation model.}
The DCCRN uses a convolutional STFT front end, complex-valued convolutional encoder-decoder blocks, a two-layer complex LSTM bottleneck, and complex ratio-mask prediction. We train it to reconstruct the retained waveform using a negative scale-invariant signal-to-noise ratio (SI-SNR) loss and AdamW. Varying the selected target subsets during training teaches the network to interpret different attenuation goals through the shared conditioning interface.

\subsubsection{Attenuation Benchmark}
We constructed evaluation mixtures covering Sona's 25 built-in sound classes using clips held out from training, following the mixture procedure in Section~\ref{sec:sona-training}. Each mixture combined designated suppression targets with background recordings from home, public indoor (e.g., coffee shop), and outdoor urban environments. We evaluated one, two, and three simultaneous targets, matching the interface's supported range. The reference signal was the intended retained soundscape: the sum of all sources not designated for suppression. We report Input SI-SNR between the unprocessed mixture and this reference, Output SI-SNR between the processed output and the same reference, and SI-SNRi, calculated as Output SI-SNR minus Input SI-SNR. These metrics characterize the reconstruction of the retained soundscape.

\begin{table}[t]
    \centering
    \caption{Attenuation benchmark results: mean SI-SNR (dB) between the retained-scene reference and the unprocessed mixture (Input), the processed output (Output), and their difference (SI-SNRi), for one to three simultaneous targets.}
    \label{tab:att_quality}
    \small
    \begin{tabular}{lccc}
        \toprule
        \multicolumn{4}{l}{\textbf{Attenuation Quality}} \\
        \midrule
        Targets & Input SI-SNR & Output SI-SNR & SI-SNRi \\
        \cmidrule(lr){2-4}
        1 & 5.01 & 8.30 & 3.29 \\
        2 & 1.54 & 4.54 & 3.00 \\
        3 & $-0.37$ & 2.86 & 3.23 \\
        \bottomrule
    \end{tabular}
\end{table}

\begin{table*}[t]
    \centering
    \caption{Runtime performance of Sona across nine consecutive 10-minute runs. H, P, and C denote the home, public indoor, and cityscape scenes described in Section~\ref{sec:sona-mobile}; the numeral gives the number of simultaneous targets (1–3).}
    \label{tab:runtime_eval}
    \begin{tabular}{lccccccccc}
    \toprule
    \textbf{Metric} & \textbf{H1} & \textbf{H2} & \textbf{H3} & \textbf{P1} & \textbf{P2} & \textbf{P3} & \textbf{C1} & \textbf{C2} & \textbf{C3} \\
    \midrule
    CPU Usage (\%) & 17.88 & 17.76 & 18.08 & 18.69 & 23.08 & 23.05 & 17.34 & 18.08 & 17.55 \\
    Inference Time (ms) & 11.32 & 11.30 & 11.34 & 10.97 & 10.98 & 11.15 & 11.22 & 11.15 & 11.21 \\
    E2E Latency (ms) & 42.03 & 42.02 & 42.01 & 41.56 & 41.61 & 41.70 & 41.74 & 41.56 & 41.72 \\
    Battery Usage (\%/hr) & 16.60 & 16.70 & 17.20 & 17.70 & 18.60 & 17.00 & 16.50 & 16.50 & 16.60 \\
    \bottomrule
    \end{tabular}
\end{table*}

Table~\ref{tab:att_quality} reports positive mean SI-SNRi across all three target-count conditions: 3.29~dB for one target, 3.00~dB for two targets, and 3.23~dB for three targets. Mean Output SI-SNR decreased from 8.30~dB with one target to 2.86~dB with three targets, while mean Input SI-SNR also decreased, from 5.01~dB to $-0.37$~dB. Improvement over the corresponding unprocessed mixtures was therefore approximately 3~dB across the tested conditions. These results provide initial evidence that Sona improves reconstruction of the intended retained soundscape when suppressing up to three simultaneous targets in constructed mixtures.


\subsubsection{Mobile Optimization}
\label{sec:sona-mobile}
We implemented the attenuation model in PyTorch and converted it to Core ML using Core ML Tools for on-device inference on iOS. The streaming pipeline processes a 250~ms sliding window every 25~ms. To characterize runtime, we profiled Sona on the study device, an iPhone 16 Pro Max, across three acoustic scenes: home (dog barking, vacuum, hair dryer), public indoor (keyboard typing, eating sounds, baby crying), and cityscape (motorcycle, construction, ambulance siren). For each scene, we evaluated one-, two-, and three-target configurations, yielding nine 10-minute profiling runs, and measured model inference time, end-to-end latency, CPU usage, and battery drain.

Table~\ref{tab:runtime_eval} shows that Sona maintains stable, low-latency performance across all conditions. Inference time averaged 11.18~ms across all runs (range: 10.97--11.34~ms), and end-to-end latency averaged 41.77~ms (range: 41.56--42.03~ms), with negligible variation across scenes and target configurations. This indicates that latency remains consistent during continuous use and does not degrade as the number of targets increases. CPU usage remained mostly stable, and the battery consumption averaged 17.04\% per hour (range: 16.5--18.6\%), corresponding to approximately 5.4--6.1~hours of continuous use on a full charge.

\subsection{Live Mediation Layer}
The Live Mediation Layer enables users to selectively attenuate bothersome sounds in the moment (DG1). Sona captures ambient audio through the microphone, retrieves embeddings for user-selected targets, and applies the attenuation framework described in Section~\ref{sec:sona-multi-target} on-device before playing the processed audio through the user's headphones.

Users can adjust how much the selected sounds are reduced using a slider ranging from 0\% to 100\%. Upon change, Sona blends the original microphone signal $x(t)$ with the framework's output $\hat{x}(t)$:
\begin{equation}
y(t) = (1-\alpha)x(t) + \alpha\hat{x}(t),
\qquad \alpha \in [0,1].
\label{eq:sona-strength}
\end{equation}
At $\alpha=0$, the original audio passes through unchanged; at $\alpha=1$, the model applies full attenuation. Intermediate values produce a proportional blend, allowing users to soften a sound without eliminating it entirely. This gives users a practical middle ground between pass-through and full suppression for instances where they want relief while still preserving some audibility of the bothersome sound (e.g., knowing a smoke alarm occurs).

To reduce friction in configuring the system during trigger encounters (DG2), Sona uses Apple's on-device sound recognizer \cite{noauthor_sound_2025} to detect sound events. We mapped Sona's built-in sounds to the recognizer's categories. When it detects a sound that is not already selected, Sona displays a three-second transient suggestion, which the user can tap to select; otherwise, it dismisses automatically.

\subsection{Personalization Layer}
Users may encounter bothersome sounds that are not covered by Sona's current sound classes. Therefore, this layer lets users extend system coverage by creating custom sound classes with short recordings from their environments (DG3).

Users record short audio samples and assign a text label (``blender'' or ``mechanical whirring''). Because users' recordings typically contain the target sound mixed with background audio, encoding the raw clip directly would produce an embedding of the scene rather than the sound. Therefore, Sona processes each sample through AudioSep's separation backbone conditioned on the users' label \cite{liu_separate_2024}. Sona then extracts intermediate features near the bottleneck, collapses the frequency dimension, and applies attentive statistics pooling over time \cite{okabe_attentive_2018} to obtain a fixed-dimensional representation. The pass is conditioned on the label so that the bottleneck activation reflects the labeled sound rather than the entire scene. Sona then averages these representations across the recordings, allowing users to update the custom class by adding more examples over time. Since the built-in classes use the same procedure (Section~\ref{sec:sona-training}), enrolling a custom class requires no changes to the deployed attenuation model.

Manually capturing a trigger requires users to operate the interface while experiencing the sound they want to avoid. Because sound-related distress can interfere with attention, information processing, and self-regulation \cite{hicks_informing_2025, guzick_how_2025, stromberg_experiences_2022}, Sona aims to minimize the interactions needed to capture the triggers by complementing the recording interface with a pipeline that helps users capture encounters with personally relevant triggers as opportunities to curate audio samples for custom class creation.

People with sound sensitivity described trigger sounds through diverse terms, such as descriptions of source, action (e.g., ``chewing''), and perceptual qualities (e.g., sharpness, pitch, repetition) \cite{landon_qualitative_2016, edelstein_misophonia_2013}. To support all of these approaches, Sona lets users describe their personal triggers in natural language (Figure~\ref{fig:user_interface}C). During an active listening session, Sona repeatedly uses Audio Flamingo 3 \cite{goel_audio_2025}, an audio language model hosted on our lab server, to describe the sounds and their perceptual qualities in the most recent five-second audio segment. Sona then uses Apple's on-device language model \cite{noauthor_foundation_2025} to compare these descriptions with the user’s sensitivity profiles and identify the best matches, if any. When it finds a match, Sona prompts the user to review and save the corresponding audio for later custom-class creation. 





\begin{figure*}[ht]
    \centering
    \includegraphics[width=\linewidth]{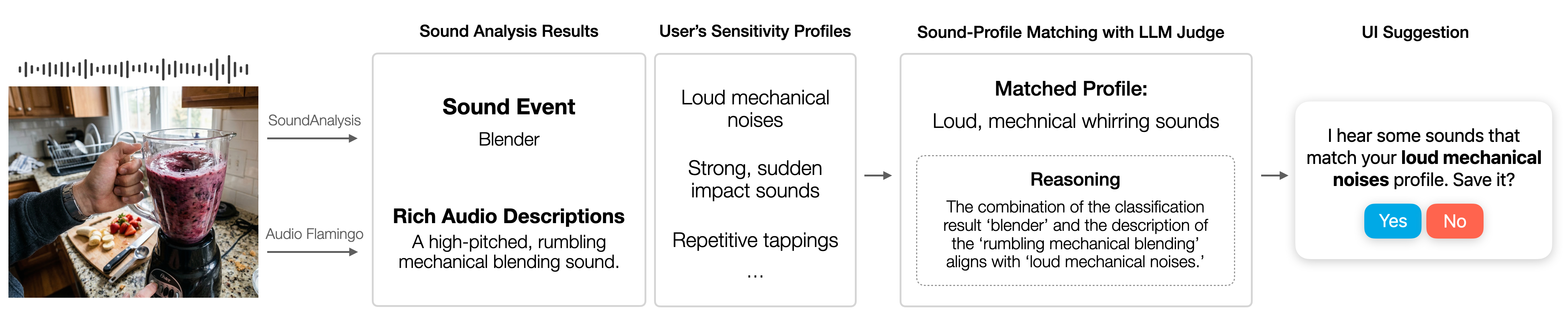}
    \caption{Sona's trigger discovery pipeline. When an unsupported sound is detected, the system generates a description and matches it against user-defined sensitivity profiles to determine whether it should be saved for later personalization.}
    \Description{Diagram of Sona's unsupported-sound discovery pipeline. An environmental sound is analyzed by Apple's SoundAnalysis and by Audio Flamingo 3, which produces a rich audio description. The classification result and description are compared against the user's stored sensitivity profiles by an on-device LLM judge. If a profile matches, the interface presents a lightweight save-or-dismiss suggestion.}
    \label{fig:unsupported_sound_discovery}
\end{figure*}

\section{User Study}
We conducted an \textit{in-situ} evaluation study with 10 participants who experience sound sensitivity to understand and gather feedback on their experience using Sona across a blend of realistic and simulated listening situations.

\subsection{Participants}
We recruited ten adults who experience sound sensitivity via email, social media, and snowball sampling (Table 4). Participants ranged from 19 to 51 years old ($M=36.4$, $SD=9.48$). As part of screening, participants completed the SSSQ-2 \cite{aazh_sound_2024}, which we used to characterize sound-sensitivity symptoms. Because SSSQ-2 does not specify a validated score, we did not apply a fixed numerical threshold for eligibility. To be eligible, participants must have indicated at least one sound-sensitivity symptom occurring on two or more days in the preceding two weeks. From this pool, we recruited participants with a range of SSSQ-2 scores and reported sound-sensitivity experiences.

\subsection{Study Apparatus}
Sona ran on an iPhone 16 Pro Max, using an external wired microphone (SonicPresence SP15C) for audio capture and Sony WH-1000XM4 headphones in wired mode for playback. Microphone placement was kept constant across sessions. When target sounds did not naturally occur, we introduced them via controlled playback on a laptop or tablet. Participants completed ratings on electronic forms. At the start of each session, we clarified that Sona modifies the audio delivered through the headphones rather than removing sound from the room itself, and asked participants to evaluate changes in prominence rather than complete removal.

\subsection{Procedure}
The user study sessions took place on a university campus and lasted approximately two hours. The session consisted of four parts: an initial interview and system walkthrough, two \textit{in-situ} listening scenarios, a personalization activity, and a concluding semi-structured interview. Our institution's Institutional Review Board (IRB) determined that the study qualified for Exemption 3 status.

\begin{table*}[t]
  \centering
  \caption[Participant demographics, SSSQ-2 symptom profiles, and
  coping strategies.]{Participant demographics, SSSQ-2 symptom
  profiles, and current coping strategies. Profiles list forms of
  sound sensitivity with scores above zero; parenthetical values show
  the obtained score and maximum possible score. Hyperacusis
  (0--12) encompasses loudness intolerance, sound-induced pain or
  discomfort, and fear of worsening hearing or tinnitus. Misophonia
  (0--3) reflects anger or anxiety elicited by specific trigger sounds,
  while noise sensitivity (0--3) reflects disturbance from
  environmental noise. Scores represent self-reported symptom
  frequency during the preceding two weeks rather than clinical
  diagnoses. A score of zero, corresponding to the 0--1 days response
  category, is omitted.}
  \label{tab:participants}
  \begingroup
  \small
  \setlength{\tabcolsep}{4pt}
  \renewcommand{\arraystretch}{1.12}
  \begin{tabular}{@{}llc
      >{\raggedright\arraybackslash}p{0.45\textwidth}
      >{\raggedright\arraybackslash}p{0.30\textwidth}@{}}
    \toprule
    \textbf{ID} &
    \textbf{Gender} &
    \textbf{Age} &
    \textbf{SSSQ-2 Symptom Profile (Score/Maximum)} &
    \textbf{Current Coping Strategies} \\
    \midrule
    P1  & Woman      & 51 &
    Hyperacusis (12/12); misophonia (3/3); noise sensitivity (3/3) &
    Avoidance \\
    P2  & Woman      & 24 &
    Hyperacusis (5/12); noise sensitivity (2/3) &
    Noise-canceling headphones \\
    P3  & Man        & 19 &
    Hyperacusis (2/12); misophonia (1/3); noise sensitivity (1/3) &
    Noise-canceling headphones, white noise \\
    P4  & Non-binary & 41 &
    Hyperacusis (3/12); misophonia (2/3); noise sensitivity (2/3) &
    Earplugs, taking breaks \\
    P5  & Man        & 46 &
    Hyperacusis (5/12); noise sensitivity (3/3) &
    Noise-canceling headphones, earplugs \\
    P6  & Woman      & 33 &
    Hyperacusis (3/12); misophonia (1/3); noise sensitivity (1/3) &
    Earplugs, noise-canceling headphones \\
    P7  & Woman      & 38 &
    Hyperacusis (10/12); misophonia (3/3); noise sensitivity (2/3) &
    Earplugs, noise-canceling headphones \\
    P8  & Woman      & 39 &
    Hyperacusis (4/12); misophonia (3/3); noise sensitivity (3/3) &
    Earplugs, breathing, avoidance \\
    P9  & Woman      & 35 &
    Hyperacusis (3/12); misophonia (3/3); noise sensitivity (1/3) &
    Closing windows, avoidance \\
    P10 & Woman      & 38 &
    Hyperacusis (2/12); misophonia (2/3); noise sensitivity (2/3) &
    Earplugs, noise-canceling headphones \\
    \bottomrule
  \end{tabular}
  \endgroup
\end{table*}

\subsubsection{Initial Interview and System Walkthrough}
We began by asking participants about their experiences with everyday sound, including sounds they found difficult and the current strategies or technologies they used to manage their sound environments. We also asked whether they preferred to avoid any sounds during the study.

We then demonstrated Sona in the lab and explained its UI and controls. Participants learned how to start a listening session, select target sounds, and adjust attenuation strength sliders. We demonstrated selected attenuation and customization features using the examples of dog barking and keyboard (targets) occurring alongside speech and knocking (non-targets). We also introduced Sona's sound recognition-based suggestions.

\subsubsection{In-Situ Use Across Two Sound Environments}
Participants used Sona in two campus environments with different soundscape characteristics: a study lounge and a busy food court. 

Participants first moved to a study lounge that was relatively quiet but still had ambient sounds. We asked them to use Sona as if they were trying to focus while doing light reading for 10 minutes. While the environment provided ambient sounds such as HVAC noise, nearby speech, and footsteps, we intermittently introduced additional sounds that could plausibly occur in such a setting, including keyboard typing, dog barking, vacuum, lawn mower, and fireworks. We encouraged participants to verbalize their thoughts while using the system.

Participants then moved to a busy food court, where we asked them to use Sona while planning a to-do list for the week on a notepad for approximately 10 minutes. This environment contained a denser, more dynamic mix of natural sounds, including nearby conversations, laughter, food preparation, television, and furniture movements like dragging chairs. In addition, we produced sounds common to public dining spaces, such as baby crying, eating-related sounds (e.g., lip smacking), and emergency sirens. As in the first scenario, we encouraged them to comment on what they noticed.

After each activity, participants completed a short questionnaire in which we asked about Sona's ability to dampen trigger sounds while retaining other sounds. They also rated their experience on 7-point scales, including two items adapted from UMUX-Lite \cite{lewis_umux-lite_2013} assessing whether Sona met their needs and whether the system was easy to use.


\subsubsection{Creating Custom Sound Classes}
After the two \textit{in-situ} activities, participants returned to the lab for the customization tasks. We conducted this part in a controlled setting so that participant-relevant target sounds could be staged repeatedly and participants could experience the full customization workflow within the session. During the session, each participant was instructed to identify one or two sounds that were personally relevant or not represented in Sona's built-in sound set. Participants then used both manual recording and audio LM-assisted capture to collect examples of a target sound played by the researcher. They submitted the recordings for processing and tested the resulting custom class by using it to attenuate the target sound. Afterward, participants rated the usefulness and ease of use of the personalization workflow using the same 7-point items.

\subsubsection{Post-Study Interview}
After completing the \textit{in-situ} scenarios and customization tasks, we conducted a semi-structured interview about participants’ experiences using Sona across the study. We asked about their overall experience; perceptions of Sona’s controls and the mediated soundscape, including when attenuation felt helpful or insufficient; experiences with the customization workflow and sound recognizer-initiated suggestions; anticipated fit with their everyday sound-management practices; and suggestions for future design.

\subsection{Data Analysis}
The first and second authors analyzed participants' comments during use and the post-study interviews using a qualitative content analysis (QCA) approach \cite{graneheim_qualitative_2004}. We segmented the transcripts into meaning units ($n$=549), condensed each, and generated codes covering the experience of the soundscape, interaction with Sona's controls, customization, anticipated everyday use, and future design suggestions. We refined these codes iteratively through synchronous discussions and collaborative editing on a shared running document. The first author then grouped the codes into categories, capturing recurring patterns and contrasts across participants. Then, both first and second authors returned to the transcripts to verify these categories. The authors resolved disagreements throughout the process through synchronous (e.g., video calls) and asynchronous (e.g., Slack) discussions.

\subsection{Findings}

\begin{figure*}[ht]
    \vspace{-1em}
    \centering
    \includegraphics[width=\linewidth]{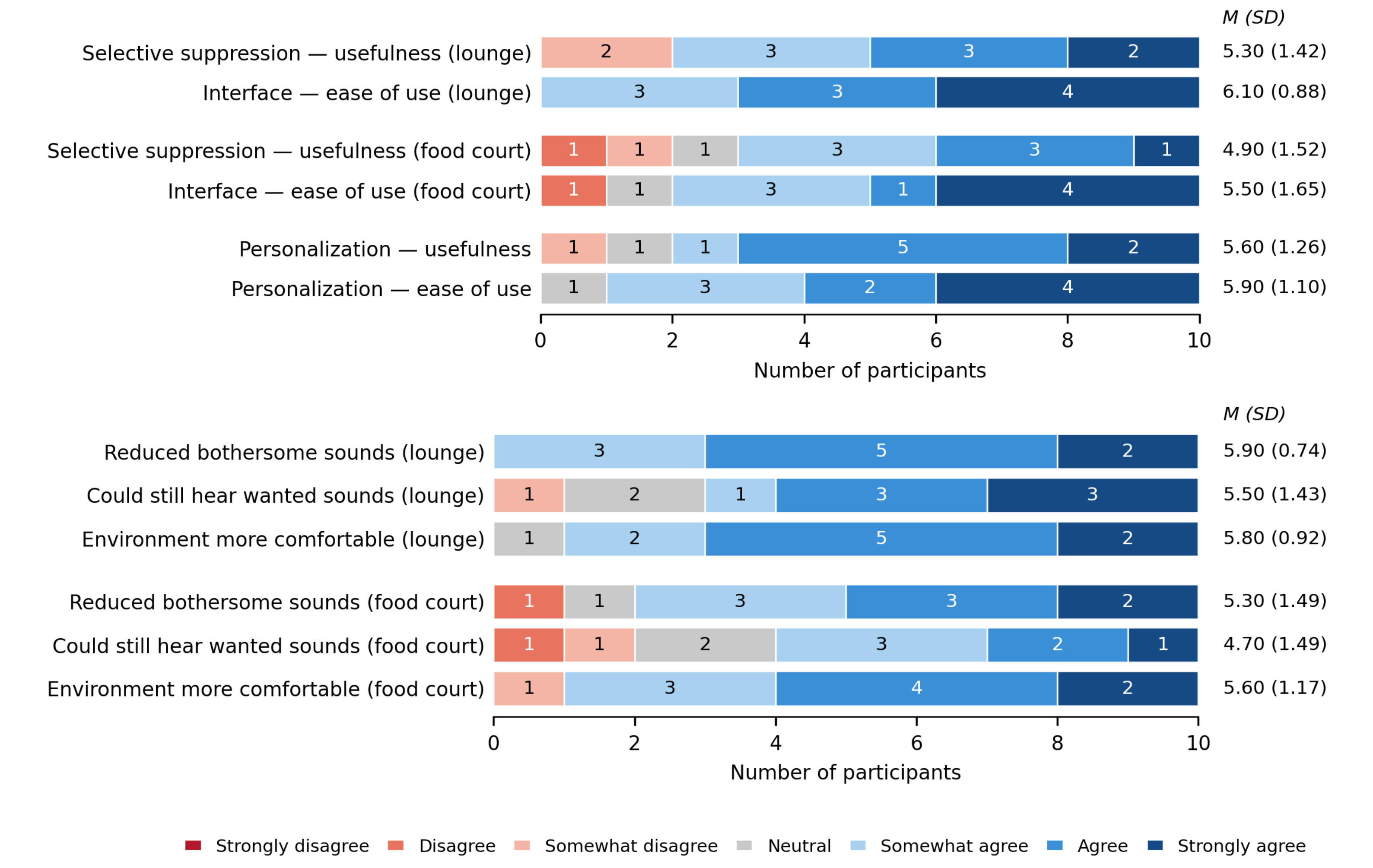}
    \vspace{-1em}
    \caption{Participant ratings ($n = 10$) on 7-point agreement scales (1 = strongly disagree, 7 = strongly agree). Stacked bars show response counts, with means ($M$) and standard deviations ($SD$) on the right. \textbf{Top}: Perceived usefulness of selective suppression and ease of using the interface, rated after each in-situ scene, and usefulness and ease of use of the Personalization Layer, rated after the personalization activity. Each measure used a single item adapted from UMUX-Lite~\cite{lewis_umux-lite_2013}. \textbf{Bottom}: Experience ratings collected after each scene for three statements, shown in shortened form in the figure: ``Sona reduced the sounds that were bothering me,'' ``I could still hear the sounds I wanted or needed to hear,'' and ``Sona made the sound environment more comfortable.''}
    \Description{Two panels of horizontal stacked bars show ratings from 10 participants on a seven-point scale from strongly disagree to strongly agree. Segments show response counts, with means and standard deviations beside each bar. The top panel shows perceived usefulness and ease of use. Mean usefulness ratings for selective suppression are 5.30 in the lounge and 4.90 in the food court; interface ease-of-use ratings are 6.10 and 5.50, respectively. Personalization receives mean ratings of 5.60 for usefulness and 5.90 for ease of use. The bottom panel shows three experience ratings for each setting. Mean ratings in the lounge and food court, respectively, are 5.90 and 5.30 for reducing bothersome sounds, 5.50 and 4.70 for hearing wanted or needed sounds, and 5.80 and 5.60 for making the environment more comfortable. Responses generally favor agreement, with higher average ratings in the lounge across all scene-specific items. Hearing wanted or needed sounds in the food court receives the lowest average experience rating and the most neutral or disagreeing responses.}
    
    \label{fig:quantitative}
    \vspace{-1em}
\end{figure*}

\subsubsection{Experience of Sona's Selective Attenuation Across Scenarios}
Participants generally found that Sona helped make their soundscapes more comfortable (lounge: $M = 5.80$, $SD = 0.92$; food court: $M = 5.60$, $SD = 1.17$) and linked the utility of Sona's selective attenuation to both reducing bothersome sounds and hearing other sounds. Most participants noticed that Sona reduced bothersome sounds in both the lounge (10/10) and food court (8/10). P8 described the immediate benefit during the reading task:
\begin{quote}
    ``It muffled the keyboard clacking. The lawnmower went `vroom,' and then it was gone. It worked really, really well.'' (P8)
\end{quote}
Being able to hear other auditory cues was part of this benefit. For example, P4 described valuing that Sona ``retained talking sounds while cutting out a lot of noise,'' and P9 similarly highlighted the preservation of knocking sounds and speech. For P9, maintaining awareness was a condition of willingness to use such a system in public: ``I would never use a headphone in public if it left me unaware of what was happening around me.''
 
However, participants described attenuation quality as inconsistent. Some selected sounds remained louder than participants preferred. For example, P2 described that even though Sona suppressed most dog barking well, it did not quiet higher-pitched barks enough. Likewise, P4 estimated that baby crying in the food court was reduced by only a quarter. Other participants reported that although speech was mostly audible, processing made it somewhat difficult to comprehend. For example, in the food court, P10 stated: ``I can still hear speech; it's just less loud. Some words I can't really make out as much unless people get louder.'' P3 similarly reported hearing the researcher speak without understanding the words, particularly when attenuation exceeded approximately 80\%. Because participants wanted to talk to nearby people while using Sona (e.g., P3: studying with friends; P1: accompanying someone in the emergency room), these findings suggested that better speech preservation is an important next step.

Attenuation also changed which sounds drew participants' attention. In the food court, P6 stated that after selecting three targets, Sona ``amplified the sound of pages turning, and that started being distracting.'' P2 and P7 likewise noticed increased prominence of ambient sounds, including laptop closing and garbage-bag rustling.

\subsubsection{Perceived Agency and System Usability}
Participants described Sona's controls as easy to use:
\begin{quote}
    ``I really like that there isn't much going on in the interface\ldots{} Here, the questions are simply: What are we blocking, and how much are we blocking it?'' (P9)
\end{quote}
Moreover, they appreciated that Sona provided a sense of control over their soundscape. This control included the ability to ``select what they want to tune out while remaining aware'' (P8, P9) and adjust the audibility of selected sounds (P1). 



Participants valued the system's ability to recognize sounds in their environment and guide them in selecting sounds to filter. P6 particularly appreciated this support when the source of a bothersome sound was unclear: ``The recognition was very helpful because sometimes it is difficult to identify what a sound is; I only know that it is bothering me.'' Although participants reported that the recognizer sometimes mislabeled sounds, they often still found the suggestions helpful. For example, in the food court, P8 accepted a suggestion recognizing a floor fan as a train because selecting the suggested sound produced the desired effect: ``When the app recognized it as a train, and I selected it, the hum turned off.'' Similarly, after a vacuum sound was labeled as an insect buzz, P6 indicated that the incorrect label would be acceptable if the selected filter still effectively suppressed the sound. These experiences suggest that, at times, participants prioritized the system's functional usefulness in recognizing a sound over the accuracy of the detected label, as it still helped them identify an effective filter in that instance.

However, participants reported that configuring the system could compete with the ongoing activities it was intended to support. While P5 agreed that the controls are easy to operate, he stated: ``When a bunch of different sounds came up, that was a little distracting. It wasn't hard, just feels like an interruption to have to turn another sound down.'' P7 similarly explained that ``having to figure out which filter to switch is distracting.'' P8 described how this interaction could extend the disruption caused by a trigger sound, sharing that, ``When a sound distracts me, and I have to open the app and look for it, that creates a longer distraction that can completely shift my focus.'' These accounts suggest that ease of configuration did not ensure that controlling the soundscape fit smoothly into an ongoing activity. To address this tension, P9 proposed separating configuration from the ongoing task: ``Doing both at once wasn't realistic for me. I could use the app, get it set to a place that felt good, and then plan my to-dos.''


On the other hand, participants suggested changes to the current controls to allow greater flexibility in how sounds were identified, attenuated, and automatically managed. Rather than applying the same attenuation strength to all selected sounds, P5 wanted to independently control the attenuation strengths for each selected sound, explaining that some air-conditioning noise could remain useful while a loud motor warranted greater suppression:
\begin{quote}
    ``I don't mind the air conditioner because it's a little bit of white noise, but I don't want it to be so loud\ldots{} Something like a loud motor, I'd say, No, cut that out.''
\end{quote}
P5 also suggested a broader sound category that covers a range of sounds that did not require him to select a specific sound source, suggesting ``more generic options, like rhythmic noises or hums.'' As he explained, ``You could just say, there's a humming noise, I don't know what it is, turn it down.'' Together, these suggestions reflect a desire for configurable filtering around the perceptual qualities and effects of a sound, rather than the predefined sound sources in the system.

Participants also considered automation as an approach for reducing repeated interactions. P8 wanted the system to learn from previous interactions and apply familiar filtering choices automatically:
\begin{quote}
    ``If a recurring sound is one that I use often, the app could shut it off automatically instead of making me select it again.''
\end{quote}
P5 similarly saw value in automatically filtering sounds already designated as unwanted, so that repeated manual toggling would become unnecessary. However, participants varied in how much control they wanted to delegate to the system. P9 described herself as ``somewhat of an AI skeptic'' and valued deliberate configuration: ``I actually prefer that this system requires me to set things up. I like having control and making the system respond to me.'' As such, while automation could alleviate user interaction burdens in managing recurring sounds, participants' responses indicate a need for configurable levels of automation that preserves individual control over when and how automation decisions are carried out.

\subsubsection{Perceptions of Personalization Workflow and Results}
Participants' perceptions of the personalization results were mixed. While P3 reported that the blender sound was no longer audible after training a custom \href{apita}{\href{}{``blender''}} class with two different samples from YouTube, most participants (N=6) reported only partial reduction of the target sounds. P4 described the ``children screaming'' class ``reduced by about 80 percent.'' P6 created a ``pen-tapping'' class and described that it ``worked very well.'' She added: ``it wasn't as effective as the built-in sounds, but it does provide some suppression.'' P10 reported less consistent attenuation when testing a custom ``arguing'' class: ``It got less loud, but I could still hear it for the most part.'' P2 experienced similar limitations with a car horn, but perceived improvements with increasing the number of samples: ``As I train it with more samples, it does better at filtering the car horn without losing your speech.'' Similarly, P8 perceived minimal effect with one sample of sniffing sound, but after adding three more samples, she reported, ``sniffing sound was gone, but I did hear the door knocking.'' 

Despite the uneven results, most participants agreed that Sona's personalization component could meet their sound management needs after experiencing the customization workflow ($M=5.60$, $SD=1.26$). P6 explained: ``Many people have very subjective sound sensitivities, so it's a very useful feature.'' P9 described how customization could benefit her routines: ``I could record the sound of children making more noise than I wanted instead of having to change my working hours.'' P3 described the customization as essential: ``Since there are a limited number of built-in sounds, it is necessary.'' 

Although participants generally considered the personalization workflow easy to use ($M=5.90, SD=1.10$), participants suggested several improvements. P9 suggested making the improvement from recordings more interpretable and visible so she could track progress. P3 suggested that the app should explicitly tell users how to provide effective samples to produce the best results. P6 desired a more efficient way to capture unexpected sounds for customization:
\begin{quote}
    ``I'm thinking about unexpected sounds. If I don't expect a sound, I probably don't have the app open... I don't want to go through 10 clicks. I want to capture it immediately.''
\end{quote}
To address this, P6 suggested embedding a recording widget in easy-to-access UIs (e.g., iOS's Control Center or lock screen).

In terms of the profile-matching workflow for capturing trigger sounds, P2 considered this the ``key in real-world use case.'' P6 agreed, explaining: ``If I'm studying, I can see how the feature would be useful.'' However, several matches were missed across sessions. For example, P9 described a ``low hum, like AC'' but Sona did not respond despite the HVAC unit being on. P4 entered ``chatters'' as a personal trigger, but Sona did not capture the sounds of a group of people having group meetings. Moreover, P5 expressed concern that Sona might misinterpret his descriptions because he ``didn't know what the system was doing behind the scenes,'' and suggested a more structured ``syntax'' for describing personal triggers rather than free text. Sharing a similar concern about transparency, P8 wanted Sona to consistently show the generated descriptions of live soundscape so she could ``assess whether it represented the intended sound.''

\subsubsection{Anticipated Usage Scenarios, Concerns, and Suggested Improvements}
Participants envisioned using Sona in a range of scenarios that contain both important auditory cues and trigger sounds. For example, P3 imagined studying alongside friends: ``It would be nice to filter what they are doing and focus on my own work while still being able to hear what they're talking about.'' P8 named a coffee shop or public library, where she could ``tune out distracting sounds while remaining focused and aware of my environment,'' and her child's swim lessons: ``I still want to be aware and have a conversation, but I might want to reduce the children screaming.'' P1 described how Sona could help her experience of accompanying a family member in the emergency room:
\begin{quote}
    ``There are all kinds of noise, people coughing and shouting. Most of the sounds in the emergency room are overwhelming, like a woman right across from me was crying and asking for some kind of pain pill. But I want to make sure I can still hear when they call me or come to me.''
\end{quote}
Five participants positioned Sona as a helpful tool for preparing them for upcoming triggering situations. For example, P1 mentioned using it for transit and stated: ``I'd set it up in advance and leave it on until I get to my destination.'' P9 mentioned configuring Sona ``before the ground crew comes every Tuesday at 2 p.m. with lawnmowers and leaf blowers.'' And P7: ``If the construction was going to continue for a week or two, I could set it up once and then be ready for those two weeks.'' While P10 agreed, she recognized that ``trying to anticipate every sound, such as glass clinking in a restaurant, can itself become substantial work.''


However, some participants were more skeptical about using Sona in environments where sounds were unpredictable, and about trigger sounds that also carried information they needed. P9 raised the former concern: ``I wouldn't use it in an unfamiliar, crowded, loud space because I don't know which sounds will appear, and the setup takes time.'' P5 drew the same line: ``I'd be less likely to use it in a rapidly changing real-world situation than in one where I already know I don't like a sound.'' P3 raised the second concern for transit:
\begin{quote}
    ``I could tune out noises like the bus door opening and closing as a custom sound, but filtering it might make me lose track of how many times it opened and therefore where I was.''
\end{quote}
While P6 preferred Sona's approach to her earplug's ``all-or-nothing,'' she raised an important concern about safety: ``I am female, and I don't feel comfortable suppressing sounds while walking down the street. Many women might not want to use it in public unless they are in a safe space.'' P8 raised a similar concern and proposed that Sona should include a ``never cancel'' list, in which it learns crucial cues like a child crying and calling for a parent because she needs to stay vigilant.

Participants raised privacy concerns about how Sona processes recordings and wanted control over when audio is captured and uploaded. P1 sought reassurance about security and data use and appreciated that the attenuation pipeline was mostly on device. P6 pointed to Sona's profile-matching pipeline and noted that she ``did not want the app listening all the time.'' After P5 learned that the recordings were sent to our lab server for custom class creation, he wanted to host the pipeline on his own desktop computer. Sona's Personalization Layer also raised concerns about recording and selectively filtering other people. Although P6 expressed little general concern about privacy, she identified sharing recordings of a particular person's voice as a potential issue. P9 stated a similar ethical concern about using Sona in social situations:
\begin{quote}
    ``I suppose you could surreptitiously record someone you didn't like and tell the app to block them, but that would raise an ethical problem.''
\end{quote}

These accounts surfaced participants' concerns about where audio was processed, when it was captured, and how those recordings were used. Participants also considered the Personalization Layer from the standpoint of others in their environment: control over one's own listening experience raised ethical questions when it involved recording and selectively filtering another person without their knowledge.

\section{Discussion}
To our knowledge, this is the first study to explore how people with sound sensitivity use and personalize selective attenuation to manage bothersome sounds in everyday settings. Participants valued the selective control of the live soundscape and the ability to extend the system. However, they also reported uneven attenuation, interaction demands, and uncertainty about the personalization process. Below, we contextualize these findings in relation to prior work and discuss implications for the future of soundscape mediation tools for people with sound sensitivity.

\subsection{Designing for the Residual Soundscape}
Participants saw the quality of the residual soundscape as an important consideration for adopting Sona in everyday life. While Sona's design partially supports this goal by reconstructing the residual acoustic scene, our findings further suggest that the qualities of this retained soundscape should be an explicit design target. Below, we describe two considerations surfaced from the user evaluation.



First, selective attenuation shifted the balance among perceived environmental sounds. Prior work suggested that perceived soundscape quality depends on the combination and relative levels of constituent sounds \cite{hong_effects_2020}. Our findings extended this insight to selective attenuation for people with sound sensitivity: participants reported that reducing selected triggers made otherwise ambient sounds, such as page turning, laptop closing, and garbage bag rustling, become more prominent and distracting. This shift can arise from the attenuation mechanism itself: removing energy from the mixture can raise the relative level of what remains. Future work should explore shaping the retained soundscape by pairing selective attenuation with user-controlled ambient sounds or light masking.

Second, trigger sounds can also carry important awareness- or task-related cues. For example, reducing bus door sounds (P3) could make it harder to track how many stops had passed. Sona's graduated control allows partial reduction of trigger sounds rather than complete removal. Future studies should explore more diverse audio transformation techniques that optimize for both sensory relief and awareness of necessary cues.

\subsection{Supporting Anticipatory and In-the-Moment Mediation}
Participants' experience controlling Sona revealed a gap between processing sound in real time and being able to configure useful mediation during an ongoing task. While participants found Sona's interface easy to use and the recognizer-initiated suggestions a helpful shortcut, selecting and assessing filters still pulled P5 and P8 out of their tasks, and in more dynamic soundscapes, the suggestions themselves became something P7 had to monitor. This finding aligns with the tension between supporting awareness of sensory triggers and the added distractions of technology identified by prior work \cite{hicks_informing_2025}. On the other hand, participants envisioned using Sona as a preparation tool for familiar situations, such as scheduled lawn maintenance (P9) and construction (P7). This reflects prior findings that describe advance planning as a key strategy for people with sound sensitivity to manage trigger exposure \cite{dibb_qualitative_2026, dotch_understanding_2023}. However, advance preparation redistributes, rather than reducing, the effort needed. As P10 put it, anticipating triggers can become stressful itself.

One way to connect preparation effort with in-the-moment controls is to let configurations carry forward. For example, Sona could associate the acoustic context with an applied configuration, and surface it again when a similar context occurs. Hearing aids already implemented a version of this mechanism by classifying the acoustic scene and switching listening programs automatically \cite{noauthor_widex_2026, noauthor_ai_2026, searchfield_performance_2018}. Research with hearing device users showed differing preferences for automatic adaptation and a desire for assistance that preserves their control over device settings \cite{luengen_connected_2021, banerjee_hearing_2011}. Our participants expressed a similar split in automation preferences: P5 and P8 wanted recurring sounds handled without repeated selection, whereas P9 valued deliberately configuring Sona and having it respond to her choices. Future soundscape mediation systems can accommodate these preferences by presenting retrieved relevant configurations as suggestions through transient prompts like those Sona uses for detected sounds, while letting users specify which configurations to apply automatically.

\subsection{Supporting Diverse Mediation Intents}
Across sessions, participants mentally mapped the desired changes to Sona's source-based sound classes. For example, P7 tried to use ``screaming'' to address the ``unpredictable cadence and changing volume'' of the overall soundscape, but heard no improvement. These experiences revealed a challenge: source-based controls forced participants to translate a recognizable bothersome quality into a unit of control that may not produce the intended effect.

P5 pointed out this mismatch more directly and asked for the ability to instruct Sona to turn down ``hum'' sounds without knowing their source. This is consistent with prior literature showing that people with sound sensitivity---including autistic adults and people who experience misophonia---described trigger sounds in a diverse way, including sources, actions, and perceptual qualities (e.g., sharpness, repetition) \cite{landon_qualitative_2016, swedo_consensus_2022, clonan_identifying_2026}. Sona steps toward supporting these richer forms of mediation intents by accepting natural-language descriptions of triggers (``mechanical whirring sounds'') to identify candidate recordings, but it still creates mostly source- (e.g., my roommate's blender) or action-based (e.g., sniffing) custom classes. Closing this gap would require effective target representations that capture perceptual qualities shared across different sound sources. Furthermore, future work should explore supporting open-vocabulary queries---such as perceptual qualities---as explicit, direct targets of real-time target sound separation, allowing PWSS to specify bothersome sounds even when they cannot identify them.

\subsection{Enabling Guided, Interpretable, and Accessible Personalization}
When creating custom classes, participants described uncertainty about the quality of their recordings and whether their personalization efforts improved attenuation. Research on personalized sound recognition similarly identified challenges in collecting and assessing representative audio recordings \cite{goodman_toward_2021}. For sound-sensitive people, this challenge is higher-stakes because these samples involve attending to sounds they want relief from. SPECTRA offers concrete approaches to supporting this work through visualizations and annotations that help users inspect recordings and compare examples \cite{goodman_spectra_2025}. Incorporating similar support into Sona could help users judge whether a recording captures the intended sound and what it contributes to a custom class. Future work can also draw on PWSS's existing support networks \cite{bennett_interdependence_2018} by inviting trusted companions to help capture and review samples.

Assessing the benefit of personalization introduces further challenges. The amount of attenuation that provides meaningful relief may differ across listeners, and reducing a trigger must be considered alongside changes to sounds they want to retain. Moreover, this process can require repeated exposure to the very sounds users wish to reduce. Therefore, a key challenge is making personalization outcomes more visible while minimizing the effort and trigger exposure required to make that judgment.


\subsection{Privacy and Societal Concerns}
Privacy considerations informed Sona’s architecture. While the Live Mediation Layer runs entirely on device, the Personalization Layer uses AudioSep \cite{liu_separate_2024} and Audio Flamingo 3 \cite{goel_audio_2025}, whose computational demands led us to host them on our lab’s server. We deliberately avoided commercial inference APIs, such as Gemini \cite{noauthor_models_2026}, to approximate a future in which personalization models can be self-hosted (also proposed by P5) or run entirely locally. However, processing location is only part of the privacy concerns: P6 preferred deliberately initiated listening sessions, while P9 wanted control over whether recordings were uploaded. These concerns suggest that future systems should provide separate, understandable controls over when audio is analyzed, transmitted, and retained.

Participants envisioned using Sona to study with friends (P3), attend a child's activities (P8), and accompany loved ones in noisy settings like an emergency room (P1). These scenarios suggest Sona's potential to support participation in situations that would otherwise be difficult to tolerate. Meanwhile, participants also raised concerns about Sona capturing or covertly filtering other people's voices, echoing concerns raised by prior work \cite{haas_interactive_2020}. Future studies should treat the soundscape as a shared, socially negotiated environment \cite{oleksik_sonic_2008} by including both PWSS and people who share their environments.

At a broader societal level, personalized soundscape mediation systems must not replace efforts to foster more sensory-friendly spaces, aligning with prior work emphasizing supporting people's immediate sensory needs while addressing the environmental and societal conditions that create barriers \cite{hicks_informing_2025}. Future deployments should examine how mediation fits alongside these arrangements and whether it expands users’ choices. This would help establish how personal control can support participation while preserving shared responsibility for accessible environments.

\subsection{Limitations}
Our user study involved ten participants, each in a single 120-minute session. Therefore, it cannot establish how Sona would be experienced during sustained daily use, including how personalization practices and attenuation preferences evolve, whether interaction demands diminish with familiarity, or whether initial impressions persist beyond novelty effects. Although both listening scenarios took place in real-world environments with naturally occurring ambient sound, we supplemented them with playback to introduce common trigger sounds across participants. Therefore, these study conditions may not capture the full variability and unpredictability of everyday trigger encounters. As a proof-of-concept prototype, Sona processes audio in mono at 16 kHz, which might have limited the fidelity of the retained soundscape. The mono pipeline does not preserve interaural cues. Although participants reported being able to hear relevant surrounding sounds, our evaluation did not assess their ability to locate those sounds. Finally, we did not formally compare Sona with existing tools such as active noise-canceling headphones or earplugs. Participants' comparisons drew on prior experience and thus offer insights into perceived relative benefits and limitations, rather than evidence of comparative effectiveness.

\section{Conclusion}
We present Sona, a mobile system that lets people with sound sensitivity (PWSS) mediate their soundscapes by selectively attenuating multiple overlapping sounds at adjustable strength in real time. It also suggests targets based on ambient sound recognition, and users can personalize it by adding custom attenuation targets from short recordings without retraining models. Sona runs on-device in real-time with approximately 42\,ms of end-to-end latency. Through an in-situ evaluation with ten PWSS across two real-life settings (study lounge and food court), participants reported that Sona made their soundscapes more manageable and valued the control it gave them over which sounds were reduced. The study also found that Sona's attenuation varied across sounds and settings, and that reducing selected sounds sometimes reshaped participants' perception of the residual soundscape. Controlling the soundscape could compete with ongoing activities, and participants needed clearer guidance and feedback to assess personalization outcomes.

These findings suggest that soundscape mediation for PWSS must treat the residual soundscape as a design target, balance user control against interaction demands, and make personalization guided and interpretable. While longer-term deployments are needed to understand how these practices evolve with sustained use, this work points toward personalized auditory accessibility that expands listeners’ choices while complementing broader efforts to make shared environments more accessible.

\bibliographystyle{ACM-Reference-Format}
\bibliography{sona_references}


\appendix

\end{document}